# How improving performance may imply losing consistency in event-triggered consensus ⋆

David Meister [a], Duarte J. Antunes [b], Frank Allgöwer [a]

[a] *University of Stuttgart, Institute for Systems Theory and Automatic Control, Stuttgart, Germany.*

[b] *Eindhoven University of Technology, Eindhoven, Netherlands.*

**Abstract**

Event-triggered control is often argued to lower the average triggering rate compared to time-triggered control while still achieving a desired control goal, e.g., the same performance level. However, this property, often called consistency, cannot be taken for granted and can be hard to analyze in many settings. In particular, the performance properties of decentralized event-triggered control schemes with respect to time-triggered control remain mostly unexplored. Therefore, in this paper, we examine these performance properties for a consensus problem considering single-integrator agent dynamics, a level-triggering rule, and a complete communication graph. We consider the long-term average quadratic deviation from consensus as a performance measure. For this setting, we show that enriching the information the local controllers use improves the performance of the consensus algorithm but renders a previously consistent event-triggered control scheme inconsistent. In addition, we do so while deploying optimal control inputs which we derive for both information cases and triggering schemes. With this insight, we can furthermore explain the relationship between two seemingly contrasting consistency results from the literature.

## 1 Introduction

In event-triggered control (ETC), a state-dependent triggering condition is evaluated online, and a sampling instant is established when this condition is satisfied, i.e., an event is triggered. On the contrary, time-triggered control (TTC) determines sampling instants solely based on time, typically resulting in periodic control. ETC often aims to reduce the average sampling rate in a control loop compared to TTC while still achieving a certain control goal, such as closed-loop performance [5,9,15,17]. Reducing the sampling rate can be valuable if closing the loop in a sampling-based fashion is costly, for example, in networked control systems [14,16].

However, it is generally challenging to formally analyze the performance and sampling rate properties of ETC compared to TTC. There are early works that arrive at such results for simple settings. For a single-integrator system, impulsive control inputs, and the output variance as a performance measure, the seminal work [5] shows that ETC with a level-triggering rule outperforms TTC by a factor of three under equal average triggering rates. This observation has sparked multiple lines of research around this potential advantage of ETC. For instance, research on ETC design has focused on finding optimal triggering rules that minimize the average sampling rate while maximizing performance, e.g., [15, Paper II] extending [5], [1] building upon [7,13,22], or [25].

Instead of seeking optimality of an ETC design, a second line of research focuses on showing the *consistency* property of ETC. According to [3], an ETC scheme is consistent if the closed-loop performance is improved compared to TTC while acting at the same average triggering rate. Some works consider ETC lowering the average triggering rate compared to TTC while still maintaining a performance level [6,21], whereas others study the property of ETC improving the performance at the same average triggering rate as TTC [2,5]. However, both perspectives achieve conceptually equivalent goals [5,20]. Several works in this area focus on developing consistent

⋆ Corresponding author: D. Meister. F. Allgöwer thanks the German Research Foundation (DFG) for support of this work within grant AL316/13-2 - 285825138 and within the German Excellence Strategy under grant EXC-2075 - 390740016. D. Meister thanks the Graduate School of the Stuttgart Center for Simulation Science for supporting him.

*Email addresses:* meister@ist.uni-stuttgart.de (David Meister), d.antunes@tue.nl (Duarte J. Antunes), allgower@ist.uni-stuttgart.de (Frank Allgöwer).

ETC strategies for linear systems with quadratic performance indices [2,3,8,26]. Similarly, $\ell_2$-performance is addressed in [6]. Moreover, [21] proposes an ETC design for continuous-time LTI systems guaranteeing the same $H_\infty$-performance as the optimal periodic controller while never exceeding its average sampling rate.

While all of these works consider single-loop systems, there is currently a lack of results formally showing the performance advantages of ETC in distributed problems [24]. To obtain a deeper understanding of design rules and influence factors in the multi-agent system (MAS) domain, we aim for a theoretical analysis of ETC performance properties in a simple yet fundamental consensus setup. Extending the seminal analysis in [5], we study a single-integrator consensus framework with quadratic performance measure, a level-triggering rule, and a complete communication graph. In [18,19], we show that TTC yields a better performance than ETC for the same average triggering rate when the number of agents is sufficiently large in this setting. Note that this is in striking contrast to the single-loop results in [5] which prove a clear performance advantage of ETC with a level-triggering rule over TTC for a single-integrator system. On the contrary, [4] considers a discrete-time variant of this problem setup. We deploy an average consensus policy and compare the performance of a level-triggered ETC scheme and an optimal synchronous TTC scheme. In contrast to [18,19], this work demonstrates numerically that the proposed ETC scheme is indeed consistent. Thus, although all works consider variants of the same consensus problem, their outcomes appear to be contradictory. On the one hand, [4] leaves us with the established belief that ETC schemes can outperform TTC also in distributed settings. On the other hand, [18,19] show that in a problem variant, this performance advantage of ETC vanishes for sufficiently many agents. In order to develop a fundamental understanding of (decentralized) ETC performance, it is important to explain the relationship of these seemingly contrasting results.

In this work, our main contributions are: (a) We uncover that the crucial difference between the two settings, which leads to the seemingly contrasting consistency results, lies in the information utilized at the local controllers. Therefore, we show that the differing outcomes are not contradictory but result from the difference in locally used information. (b) We reveal that enriching the information utilized by the local controllers in a distributed problem can render a previously consistent ETC scheme inconsistent, although it improves TTC and ETC performance. (c) We present optimal control inputs for the examined performance measure under time-triggered sampling as well as sampling with symmetric event-triggering rules. Thereby, we prove that the performance of the studied ETC schemes cannot be improved with other controllers and ensure that we compare to the optimal periodic controllers in our (in)consistency statements.

Furthermore, this article provides the following technical contributions: (a) We establish a purely analytical consistency proof for the strategy proposed in [4]. (b) We generalize the periodic scheme used as a comparison baseline in [4] to an asynchronous scheme, i.e., agents are not required to exchange information simultaneously in this work. (c) We quantify the performance gain of the ETC scheme in [4], and deduce that the triggering schemes in [19] outperform the ones in [4]. This is due to enriching the locally utilized information in [4] by local information at triggering instants.

In Section 2, we present the problem setup. In Section 3, we derive the optimal control inputs for the ETC and TTC schemes considered in this work. Subsequently, in Section 4, we analyze the performance relationship of TTC and ETC with a decentralized level-triggering rule. In particular, we differentiate between two cases regarding the locally utilized information. The simulation performed in Section 5 confirms our theoretical findings. Lastly, we give concluding remarks in Section 6.

*Notation:* A communication graph $\mathcal{G} = (\mathcal{V}, \mathcal{E})$ consists of vertices $\mathcal{V} = \{1, \dots, n\}$, called nodes, and edges $\mathcal{E} \subset \mathcal{V} \times \mathcal{V}$. In this work, we focus on undirected graphs $\mathcal{G}$, i.e., $(i,j) \in \mathcal{E} \Leftrightarrow (j,i) \in \mathcal{E}$. The remaining definitions are therefore tailored to those. Two nodes $i$ and $j$ are neighbors and adjacent if $(i,j) \in \mathcal{E}$. Hence, we define adjacency matrix $A$ with $a_{ij} = 1$ if $(i,j) \in \mathcal{E}$ and $a_{ij} = 0$ otherwise. We do not consider self-loops, i.e., $(i,i) \notin \mathcal{E}$. For undirected graphs, $A$ is symmetric. Moreover, the degree $d_i$ of node $i$ denotes the number of its neighbors, yielding the degree matrix $D := \mathrm{diag}(d_1, ..., d_n)$. For undirected graphs, the graph Laplacian $L := D - A$ is positive semi-definite. A complete graph contains edges between all distinct nodes. We thus arrive at $L = nI_n - 1_n 1_n^\top$ for complete graphs, where $I_n$ is an $n \times n$ identity matrix and $1_n$ is an $n$-dimensional vector of all ones. In our work, the nodes represent agents and the edges indicate their communication abilities. Furthermore, we denote the natural numbers by $\mathbb{N}$ and $\mathbb{N}_0 := \mathbb{N} \cup \{0\}$. We refer to the Dirac delta impulse as $\delta(\cdot)$ and the Heaviside step function as $H(\cdot)$. Let $\mathbb{1}_{(\cdot)}$ denote the indicator function, returning 1 if $(\cdot)$ holds, and 0 otherwise.

We examine stopping times of the underlying stochastic process, i.e., a random time at which a random event related to the process happens (typically for the first time). This random event is described by a stopping rule, typically as a condition on the path of the process [23]. Since the triggering rules in this work are stopping rules, the notion of symmetric stopping times will be useful.

**Definition 1 (Symmetric stopping time)** *Let $W := \{w(t) : t \geq 0\}$ be a stochastic process. A stopping time $\tau$ with respect to $W$ is called symmetric if exchanging $W$ by $-W := \{-w(t) : t \geq 0\}$ does not affect the stopping time. In that case, the corresponding stopping rule is also referred to as symmetric.*

## 2 Problem formulation

In this section, we introduce the problem setup, our notation for triggering instants, and the two information scenarios considered in this work.

### *2.1 Setup*

Let us consider a multi-agent system consisting of $n$ agents with perturbed single-integrator dynamics

$$\mathrm{d}x_i = u_i \,\mathrm{d}t + \mathrm{d}v_i, \tag{1}$$

with scalar agent state $x_i(t)$ at time $t$, control input $u_i(t)$, and standard Wiener process $v_i(t)$ where the latter is independent and identically distributed (i.i.d.) for all agents $i$ in the MAS. We will sometimes refer to the stacked state, input, and noise vectors as $x(t)$, $u(t)$, and $v(t)$, respectively, where $x(t) := [x_1(t), \dots, x_n(t)]^\top$, and $u(t)$ and $v(t)$ are defined similarly.

Since we consider a complete communication graph, every agent is able to communicate directly with all other agents. The agents use their communication capabilities to share state information with the other agents in order to achieve the overarching control goal. In this paper, we will study two different schemes for triggering transmissions: time- and event-triggered schemes. In both cases, an agent broadcasts local state information whenever one of its triggering instants is reached. This information can then be used by all agents in order to compute adequate control inputs.

The control goal is to preserve consensus among the agents while they are perturbed by noise. All agents are assumed to be initialized in consensus with $x_i(0) = 0$. As a performance measure, we consider

$$J := \limsup_{M\to\infty} \frac{1}{M}\mathbb{E}\left[\int_0^M x(t)^\top L x(t)\,\mathrm{d}t\right], \tag{2}$$

which quantifies the average quadratic deviation from consensus on an infinite time horizon. Note that $x(t)^\top L x(t) = \frac{1}{2}\sum_{(i,j)\in\mathcal{E}}(x_i(t) - x_j(t))^2$. Thus, we study asymptotic (and not transient) performance, i.e., preservation of consensus instead of convergence to consensus, cf. [10,11] for ETC schemes addressing the latter.

In this article, we will compare ETC schemes for this setup with the respective optimal periodic control schemes in order to establish (in)consistency results with respect to (2).

**Definition 2 (Consistency, [4])** *An ETC scheme is called* consistent (w.r.t. a performance measure) *if it performs at least as well as the respective optimal periodic control scheme with respect to the performance measure in consideration given equal average triggering rates.*

Consistency hence allows us to evaluate if an ETC design can generate advantages compared to TTC with respect to a performance measure of interest. The focus of this paper therefore lies on the respective performance analysis and on illustrating new phenomena and their explanation in this realm. While the simple consensus setup considered in this work does not cover the large variety of problem setups and solutions presented in the literature, the simplicity of the setting is favorable for the previously described goal. In fact, it allows for analytical performance examinations and a great level of intuition. Note that the intention behind this work is to provide fundamental results on event-triggering behavior and its implications in an explainable manner. As laid out in the introduction, such results are rarely available even in simple settings such as this one. It is left for future research to generalize or transfer our findings to other setups.

### *2.2 Triggering instants*

We utilize two alternatives for denoting triggering instants. On the one hand, we refer to *local* triggering instants with $(t_k^i)_k$ defined as the triggering instants at which agent $i$ is the initiator. On the other hand, let $(t_k)_k$ denote the sequence of *global* triggering instants, i.e., all triggering instants in the MAS. We define $t_0^i = 0$ for all $i \in \mathcal{V}$ and $t_0 = 0$. These sequences are naturally related in the sense that $(t_k)_k$ consists of all triggering instants in $(t_k^i)_k$ for all agents $i$ in an increasing order.

Furthermore, we denote the latest local and global triggering instants with respect to time $t$, *including* $t$ itself, by $t^i_{\hat{k}_i(t)}$ with $\hat{k}_i(t) := \sup\{k \in \mathbb{N}_0 \mid t_k^i \le t\}$ and $t_{\hat{k}(t)}$ with $\hat{k}(t) := \sup\{k \in \mathbb{N}_0 \mid t_k \le t\}$, respectively. Moreover, we use the shorthand $t^-$ to denote the time right before a potential triggering instant at $t$. Thus, when writing $t_k^-$, we refer to the point in time right before triggering at $t_k$. Hence, we denote the latest global and local triggering instants with respect to $t$, *excluding* $t$ itself, by $t^i_{\hat{k}_i(t^-)}$ with $\hat{k}_i(t^-) := \sup\{k \in \mathbb{N}_0 \mid t_k^i < t\}$ and $t_{\hat{k}(t^-)}$ with $\hat{k}(t^-) := \sup\{k \in \mathbb{N}_0 \mid t_k < t\}$.

### *2.3 Information Scenarios*

We consider two different information scenarios.

(1) *Controller only using broadcast information*: In the first scenario, the local controllers can only utilize broadcast information

$$\mathcal{I}_t^{\mathrm{b}} := \bigcup_{i\in\mathcal{V}}\{x_i(s) \mid 0 \le s \le t^i_{\hat{k}_i(t)}\}.$$

Hence, each local controller has access to state information of every agent up to each latest local triggering instant.

(2) *Controller also using local state at triggering instants*: In the second scenario, they can *additionally* rely on the *local* state at triggering instants, hence, on the information set

$$\mathcal{I}_{t,i}^{\mathrm{bl}} := \mathcal{I}_t^{\mathrm{b}} \cup \{x_i(s) \mid 0 \leq s \leq t_{\hat{k}(t)}\}.$$

We will demonstrate in the next section that the control inputs studied in [4] and [18] can be subsumed under the optimal control inputs derived for these information scenarios, respectively.

## 3 Optimal control inputs

In this section, we identify optimal control inputs with respect to performance measure (2) among all periodic and event-triggered controllers with symmetric triggering rules. Note that we do so while conditioning on the different information sets introduced in Section 2.3. In other words, the optimal controller depends on the available information.

Before examining the different information scenarios, let us introduce the variable

$$\hat{x}_i(t; \mathcal{I}_t) := \mathbb{E}[x_i(t) \mid \mathcal{I}_t], \tag{3}$$

which can be interpreted as state estimate given the available information at time $t$. It can be computed for both information sets introduced above, replacing the placeholder $\mathcal{I}_t$ by the respective information set of interest. Along the same lines, we define $\hat{x}_i(t^-; \mathcal{I}_{t^-})$ as the state estimate right before a potential triggering instant at time $t$. In contrast to $\mathcal{I}_t$, the information set $\mathcal{I}_{t^-}$ contains information up to time $t$ *excluding* information made available at time $t$.

### 3.1 Controller only using broadcast information

In this section, we present the class of optimal control inputs under $\mathcal{I}_t^{\mathrm{b}}$. We introduce $c(\mathcal{I}_t^{\mathrm{b}}, t)$ to be a common consensus point among all agents which may depend on broadcast information and time and requires information exchange among the agents.

**Theorem 3** *Let $(t_k)_{k\in\mathbb{N}}$ be a sequence of symmetric stopping times, and $c(\mathcal{I}_{t_k}^{\mathrm{b}}, t_k)$ be a common consensus point among all agents at those triggering instants. If agents with dynamics* (1) *use information set $\mathcal{I}_t^{\mathrm{b}}$, the impulsive control input $u(t) = [u_1(t), \dots, u_n(t)]^\top$ with*

$$u_i(t) = \sum_{k\in\mathbb{N}} \delta(t - t_k)(c(\mathcal{I}_{t_k}^{\mathrm{b}}, t_k) - \hat{x}_i(t_k; \mathcal{I}_{t_k}^{\mathrm{b}})),$$

*is an optimal control input with respect to* (2)*. Moreover,* (3) *is given by $\hat{x}_i(t; \mathcal{I}_t^{\mathrm{b}}) = x_i(t)$ if agent $i$ triggers at $t$, and $\hat{x}_i(t; \mathcal{I}_t^{\mathrm{b}}) = c(\mathcal{I}_{t^-}^{\mathrm{b}}, t)$ otherwise.*

The proof can be found in Appendix A.

The consensus point is free within the class of optimal controllers derived in Theorem 3 and can be determined based on the specific application. Two examples for such controllers are

$$u_i(t) = \sum_{k\in\mathbb{N}} \delta(t - t_k)(\bar{x}(t_k) - \hat{x}_i(t_k; \mathcal{I}_{t_k}^{\mathrm{b}})), \tag{4}$$

with $\bar{x}(t) := \frac{1}{n}\sum_{i=1}^n \mathbb{E}\big[x_i(t) \mid \mathcal{I}_t^{\mathrm{b}}\big]$, and

$$u_i(t) = \sum_{j\in\mathcal{V}} \sum_{k\in\mathbb{N}} \delta(t - t_k^j)(x_j(t_k^j) - \hat{x}_i(t_k^j; \mathcal{I}_{t_k^j}^{\mathrm{b}})). \tag{5}$$

Note that (4) resembles the continuous-time analogue of the average consensus controller employed in [4]. We have thus shown that it is optimal when assuming that the local controllers only use broadcast information. Moreover, control input (5) is quite similar to the one in [18], but utilizes $\hat{x}_i(t_k^j; \mathcal{I}_{t_k^j}^{\mathrm{b}})$ instead of $x_i(t_k^j)$. We will highlight this difference in more detail in the next section.

**Remark 4** *We emphasize that $c(\mathcal{I}_t^{\mathrm{b}}, t)$ is assumed to require information exchange among agents. As usual in the consensus literature, there also exist trivial solutions to the consensus problem such as $c(\mathcal{I}_t^{\mathrm{b}}, t) = 0$. Our technical results regarding triggering rates and performance equally hold for these schemes. However, some of these strategies do not even require cooperation, rendering the goal of saving transmissions inadequate. Therefore, we presume that saving triggering instants is a meaningful goal under the considered choice of $c(\mathcal{I}_t^{\mathrm{b}}, t)$.*

### 3.2 Controller also using local state at trig. instants

In this section, we recapitulate the class of optimal control inputs from [19] which is conditioned on the respective information sets $\mathcal{I}_{t,i}^{\mathrm{bl}}$. We then provide two control input examples that are analogous to (4), (5).

We can prove that a class of optimal control inputs under information sets $\mathcal{I}_{t,i}^{\mathrm{bl}}$ emerges from the problem considering their union $\mathcal{I}_t^{\mathrm{bl}} := \bigcup_{i\in\mathcal{V}} \mathcal{I}_{t,i}^{\mathrm{bl}} = \{x(s) \mid 0 \leq s \leq t_{\hat{k}(t)}\}$ instead. Since $\mathcal{I}_{t,i}^{\mathrm{bl}} \subseteq \mathcal{I}_t^{\mathrm{bl}}$ for all $i \in \mathcal{V}$, it suffices to derive the class of optimal control inputs under $\mathcal{I}_t^{\mathrm{bl}}$ and show that there exist local controllers within that class that only utilize $\mathcal{I}_{t,i}^{\mathrm{bl}}$.

**Theorem 5 ([19])** *Let $(t_k)_{k\in\mathbb{N}}$ be a sequence of symmetric stopping times, and $c(\mathcal{I}_{t_k}^{\mathrm{b}}, t_k)$ be a common consensus point among all agents at those triggering instants. If agents $i$ with dynamics* (1) *use information* *sets $\mathcal{I}^{\mathrm{bl}}_{t,i}$, respectively, the impulsive control input $u(t) = [u_1(t), \dots, u_N(t)]^\top$ with*

$$u_i(t) = \sum_{k\in\mathbb{N}} \delta(t - t_k)(c(\mathcal{I}^{\mathrm{b}}_{t_k}, t_k) - x_i(t_k)),$$

*is an optimal control input with respect to* (2)*.*

Two examples for such controllers are

$$u_i(t) = \sum_{k\in\mathbb{N}} \delta(t - t_k)(\bar{x}(t_k) - x_i(t_k)), \tag{6}$$

again with $\bar{x}(t) := \frac{1}{n}\sum_{i=1}^{n} \mathbb{E}\big[x_i(t) \mid \mathcal{I}^{\mathrm{b}}_t\big]$, and

$$u_i(t) = \sum_{j\in\mathcal{V}}\sum_{k\in\mathbb{N}} \delta(t - t^j_k)(x_j(t^j_k) - x_i(t^j_k)). \tag{7}$$

The control input (7) is equivalent to the one considered in [18]. Hence, we can motivate the control inputs considered in [4] and [18] as optimal controllers given different assumptions on the locally utilized information. When comparing the results in this section to Theorem 3 and (4), (5), we observe that they differ in which information is used for the state estimates. One uses $\hat{x}_i(t_k; \mathcal{I}^{\mathrm{b}}_{t_k}) = \mathbb{E}\big[x_i(t_k) \mid \mathcal{I}^{\mathrm{b}}_{t_k}\big]$ whereas the other one uses actual local state information $\hat{x}_i(t_k; \mathcal{I}^{\mathrm{bl}}_{t_k,i}) = x_i(t_k)$. Thus, in both cases all agents react impulsively when receiving new information. However, the magnitude of the impulses depends on the information scenario. Next, let us examine the implications of that difference regarding consistency of the ETC schemes.

## 4 Consistency analysis

Let us perform the desired (in)consistency analysis for both information scenarios in this section. Thereby, we arrive at an explanation for the differing outcomes regarding consistency in [4] and [18]. We consider ETC schemes with level-triggering conditions of the form

$$|x_i(t) - \hat{x}_i(t^-; \mathcal{I}_{t^-})| \geq \Delta, \tag{8}$$

as examined in [4] and [18] and where $\Delta$ is a positive constant. Triggering conditions of this kind have been proposed in multiple works in the literature and have shown consistency potential in single-loop setups of a similar kind [5,15]. Note that we consider homogeneous triggering conditions among all agents as dynamics and cost contribution are homogeneous as well.

### *4.1 Controller only using broadcast information*

Let us consider the ETC scheme with triggering condition (8) and $\Delta_{\mathrm{b}}$ denoting the constant threshold. For all $i \in \mathcal{V}$, the resulting triggering instants are

$$t^i_{k+1} = \inf\{t > t^i_k \mid |x_i(t) - \hat{x}_i(t^-; \mathcal{I}^{\mathrm{b}}_{t^-})| \geq \Delta_{\mathrm{b}}\}.$$

The corresponding inter-event times $(t^i_k - t^i_{k-1})_{k\in\mathbb{N}}$ are identically distributed according to the stopping time

$$T^{\mathrm{b}}_{\mathrm{ET}}(\Delta_{\mathrm{b}}) := \inf\{t > 0 \mid |x_i(t)| \geq \Delta_{\mathrm{b}}\}. \tag{9}$$

This ETC scheme is analogous to the one proposed in [4] but in continuous time. In this section, we present an analytical consistency proof for it, including a closed-form expression for its performance advantage over TTC. Before stating our main findings, let us present the following preliminary result.

**Lemma 6** *Let the inter-event times $(t^i_k - t^i_{k-1})_{k\in\mathbb{N}}$ be independent and identically distributed and let the corresponding stopping time be symmetric and finite in expectation. Then, given the agent dynamics* (1) *and an optimal control input as per Theorem 3, the cost according to* (2) *is given by*

$$J^{\mathrm{b}}(T_1) = n(n-1)\frac{1}{\mathbb{E}[T_1]}\mathbb{E}\left[\int_0^{T_1} v_1(t)^2 \,\mathrm{d}t\right],$$

*where $T_1 = t^1_1$ denotes the stopping time for agent 1.*

The proof can be found in Appendix B.

Note that we can analyze synchronous and asynchronous schemes alike. As a comparison baseline for the ETC scheme, let us consider a potentially asynchronous periodic scheme with constant inter-event times $T^{\mathrm{b}}_{\mathrm{TT}} > 0$.

**Proposition 7** *Suppose agents with dynamics* (1) *are controlled by optimal local control inputs according to Theorem 3 with constant inter-event times $t^i_k - t^i_{k-1} = T^{\mathrm{b}}_{\mathrm{TT}}$ for all $i \in \mathcal{V}$ and $k \in \mathbb{N}$. Then, the performance measure* (2) *evaluates to*

$$J^{\mathrm{b}}_{\mathrm{TT}}(T^{\mathrm{b}}_{\mathrm{TT}}) = n(n-1)\frac{T^{\mathrm{b}}_{\mathrm{TT}}}{2}.$$

**PROOF.** Note that constant inter-event times satisfy all assumptions in Lemma 6. Evaluating directly yields

$$\begin{aligned} J^{\mathrm{b}}(T^{\mathrm{b}}_{\mathrm{TT}}) &= n(n-1)\frac{1}{T^{\mathrm{b}}_{\mathrm{TT}}}\mathbb{E}\left[\int_0^{T^{\mathrm{b}}_{\mathrm{TT}}} v_1(t)^2 \,\mathrm{d}t\right] \\ &= n(n-1)\frac{1}{T^{\mathrm{b}}_{\mathrm{TT}}}\int_0^{T^{\mathrm{b}}_{\mathrm{TT}}} t \,\mathrm{d}t = n(n-1)\frac{T^{\mathrm{b}}_{\mathrm{TT}}}{2}. \end{aligned}$$

Let us also compute the performance of the ETC scheme utilizing triggering condition (8). We abbreviate $J^{\mathrm{b}}_{\mathrm{ET}}(\Delta_{\mathrm{b}}) := J^{\mathrm{b}}(T^{\mathrm{b}}_{\mathrm{ET}}(\Delta_{\mathrm{b}}))$.

**Proposition 8** *Suppose agents with dynamics* (1) *are controlled by optimal local control inputs according to Theorem 3 with inter-event times resulting from* (9). *Then, the performance measure* (2) *evaluates to*

$$J^{\mathrm{b}}_{\mathrm{ET}}(\Delta_{\mathrm{b}}) = n(n-1)\frac{\mathbb{E}\big[T^{\mathrm{b}}_{\mathrm{ET}}(\Delta_{\mathrm{b}})\big]}{6},$$

*where* $\mathbb{E}\big[T^{\mathrm{b}}_{\mathrm{ET}}(\Delta_{\mathrm{b}})\big] = \Delta_{\mathrm{b}}^2$.

**PROOF.** The stopping time (9) with initial condition $x_i(t) = 0$ satisfies the assumptions in Lemma 6 with $\mathbb{E}\big[T^{\mathrm{b}}_{\mathrm{ET}}(\Delta_{\mathrm{b}})\big] = \Delta_{\mathrm{b}}^2$. Note that we have reformulated the initial problem to the Lebesgue Sampling case studied in [5]. Consequently, we can deduce from [12] that $J^{\mathrm{b}}_{\mathrm{ET}}(\Delta_{\mathrm{b}}) = n(n-1)\mathbb{E}\big[T^{\mathrm{b}}_{\mathrm{ET}}(\Delta_{\mathrm{b}})\big]/6$.

With Propositions 7 and 8, we can conclude this section with a consistency result on the considered ETC scheme.

**Theorem 9** *Let agents with dynamics* (1) *be controlled by optimal control inputs as per Theorem 3. Then, the ETC scheme with inter-event times resulting from* (9) *outperforms the corresponding periodic controller with respect to performance measure* (2) *under equal average inter-event times* $T^{\mathrm{b}}_{\mathrm{TT}} = \mathbb{E}\big[T^{\mathrm{b}}_{\mathrm{ET}}(\Delta_{\mathrm{b}})\big]$. *In particular, we arrive at the performance ratio*

$$\frac{J^{\mathrm{b}}_{\mathrm{ET}}(\Delta_{\mathrm{b}})}{J^{\mathrm{b}}_{\mathrm{TT}}(\mathbb{E}\big[T^{\mathrm{b}}_{\mathrm{ET}}(\Delta_{\mathrm{b}})\big])} = \frac{1}{3}.$$

*The considered ETC scheme is thus consistent with respect to performance measure* (2).

**PROOF.** Utilizing Propositions 7 and 8 together with $T^{\mathrm{b}}_{\mathrm{TT}} = \mathbb{E}\big[T^{\mathrm{b}}_{\mathrm{ET}}(\Delta_{\mathrm{b}})\big]$ directly yields the desired results.

We can therefore conclude that the cost is three times lower with the ETC scheme than with the TTC scheme for arbitrary but equal average triggering rates satisfying $T^{\mathrm{b}}_{\mathrm{TT}} = \mathbb{E}\big[T^{\mathrm{b}}_{\mathrm{ET}}(\Delta_{\mathrm{b}})\big]$. In other words, any choice of $\Delta_{\mathrm{b}}$ and controller $u_i(t)$ within the class from Theorem 3 will lead to a better performance than the corresponding periodic controller with the same average triggering rate. As in multiple works in the single-loop case [3,5,13,15], this exemplifies once more the potential of ETC in terms of triggering rate reduction while still achieving a control goal, such as maintaining a certain performance level. In addition, this result is in line with the findings in [4] but provides a purely analytical consistency proof for the generalized continuous-time version of the setting. It thereby additionally quantifies the performance advantage of this ETC scheme in a closed form and recovers the numerical results shown in [4].

### *4.2 Controller also using local state at trig. instants*

In this section, we recapitulate (in)consistency results for the case of locally utilized information $\mathcal{I}^{\mathrm{bl}}_{t,i}$ for which we have arrived at the optimal control input and respective examples in Section 3.2. We consider the same triggering condition (8) with $\Delta_{\mathrm{bl}}$ as positive constant triggering threshold. Due to the control inputs in consideration, this results in the *global* triggering instants

$$t_{k+1} = \inf\{t > t_k \mid \exists i \in \mathcal{V} : |x_i(t) - x_i(t_k)| \geq \Delta_{\mathrm{bl}}\}.$$

Recall that global triggering instants $(t_k)_k$ refer to the "union" of local triggering instants $(t^i_k)_k$, as introduced in Section 2.2. The respective inter event-times $(t_k - t_{k-1})_{k\in\mathbb{N}}$ are identically distributed according to the stopping time

$$T^{\mathrm{bl}}_{\mathrm{ET}}(\Delta_{\mathrm{bl}}) := \inf\{t > 0 \mid \exists i \in \mathcal{V} : |x_i(t)| \geq \Delta_{\mathrm{bl}}\}. \quad (10)$$

We have utilized that $\hat{x}_i(t^-; \mathcal{I}^{\mathrm{bl}}_{t^-,i}) = x_i(t_{\hat{k}(t^-)})$.

**Remark 10** *Note that we have used the* global *sequence of triggering instants* $(t_k)_k$ *in this section to define the event-triggering behavior in contrast to the* local *sequences of triggering instants* $(t^i_k)_k$ *utilized in the previous section. In both cases, a level-triggering rule only utilizing locally available information is deployed. Depending on the applied control inputs, one or the other formulation offers advantages in terms of mathematical analysis. We will provide an elaborate discussion on the relationship between the two settings in the next section.*

We have already shown in Section 3.2 that we have managed to formulate the problem under information sets $\mathcal{I}^{\mathrm{bl}}_{t,i}$ such that we arrive at the setup from [19]. Therefore, we can recapitulate the relevant (in)consistency results and, thereby, highlight that they also hold for the control input examples provided in Section 3.2.

Let us first consider the periodic case with constant inter-event times $T^{\mathrm{bl}}_{\mathrm{TT}} > 0$.

**Proposition 11 ([19])** *Suppose agents with dynamics* (1) *are controlled by an optimal input according to Theorem 5 with constant inter-event times* $T^{\mathrm{bl}}_{\mathrm{TT}}$. *Then, the cost* (2) *is given by*

$$J^{\mathrm{bl}}_{\mathrm{TT}}(T^{\mathrm{bl}}_{\mathrm{TT}}) = n(n-1)\frac{T^{\mathrm{bl}}_{\mathrm{TT}}}{2}.$$

Note that this result is analogous to the one stated in Proposition 7. We will comment on the relationship between the two results in more detail in the next section.

Abbreviating $J^{\mathrm{bl}}_{\mathrm{ET}}(\Delta_{\mathrm{bl}}) := J^{\mathrm{bl}}(T^{\mathrm{bl}}_{\mathrm{ET}}(\Delta_{\mathrm{bl}}))$, let us also recapitulate the (in)consistency result regarding the ETC scheme with controllers as per Theorem 5.

**Theorem 12 ([19])** *Suppose agents with dynamics* (1) *are controlled by an optimal input according to Theorem 5 with inter-event times resulting from* (10)*. Then, there exists an $n_0$ such that for all $n \geq n_0$, we have*

$$J_{\mathrm{ET}}^{\mathrm{bl}}(\Delta_{\mathrm{bl}}) > J_{\mathrm{TT}}^{\mathrm{bl}}(\mathbb{E}\big[T_{\mathrm{ET}}^{\mathrm{bl}}(\Delta_{\mathrm{bl}})\big]),$$

*i.e., TTC outperforms ETC for all $n \geq n_0$ under equal average triggering rates.*

We have shown that the proposed ETC scheme is not consistent under the information sets $\mathcal{I}_{t,i}^{\mathrm{bl}}$ for sufficiently many agents in the MAS. This is in contrast to the result in the previous section which leads us to relating the two settings in more detail.

### *4.3 Relating the previous results*

Firstly, the control inputs and triggering conditions deployed in both information scenarios are conceptually similar, but essentially differ in the evaluation of the state estimate $\hat{x}(t; \mathcal{I}_t)$. This also leads to different system behaviors for the two cases. Consequently, for each case, either the local or the global perspective on triggering instants can be beneficial for the mathematical analysis, i.e., $(t_k^i)_k$ versus $(t_k)_k$. Since we consider homogeneous agents, both perspectives are equally valid for (in)consistency statements on the respective ETC scheme for this MAS problem. On the contrary, when it comes to relating the two perspectives, the local average triggering rate needs to be scaled by $n$ to make it comparable to a global triggering rate. This is due to the fact that we consider a homogeneous setup such that, on average, $n$ events are triggered globally within the local average inter-event time. Thus, in order to compare the performance results from Sections 4.1 and 4.2, we require equal global average triggering rates, i.e.,

$$\frac{n}{T_{\mathrm{TT}}^{\mathrm{b}}} \overset{!}{=} \frac{n}{\mathbb{E}\big[T_{\mathrm{ET}}^{\mathrm{b}}(\Delta_{\mathrm{b}})\big]} \overset{!}{=} \frac{1}{\mathbb{E}\big[T_{\mathrm{ET}}^{\mathrm{bl}}(\Delta_{\mathrm{bl}})\big]} \overset{!}{=} \frac{1}{T_{\mathrm{TT}}^{\mathrm{bl}}}. \quad (11)$$

We can therefore relate the two periodic schemes yielding

$$\frac{J_{\mathrm{TT}}^{\mathrm{b}}(nT_{\mathrm{TT}}^{\mathrm{bl}})}{J_{\mathrm{TT}}^{\mathrm{bl}}(T_{\mathrm{TT}}^{\mathrm{bl}})} = n. \quad (12)$$

Recall that, when considering the information utilized by the agents to be $\mathcal{I}_t^{\mathrm{b}}$, we can at best utilize the optimal controllers as per Theorem 3 while enriching this information set to $\mathcal{I}_{t,i}^{\mathrm{bl}}$ allows us to deploy the optimal controllers given in Theorem 5. Therefore, (12) shows that under equal average triggering rates, the optimal periodic scheme utilizing $\mathcal{I}_{t,i}^{\mathrm{bl}}$ outperforms the optimal periodic scheme under $\mathcal{I}_t^{\mathrm{b}}$ by a factor of $n$. Consequently, as intuitively expected, utilizing more information locally improves the controller performance. We can establish a similar result numerically for the two ETC schemes. As depicted in Fig. 1, the ETC scheme based on $\mathcal{I}_{t,i}^{\mathrm{bl}}$ outperforms the one relying on $\mathcal{I}_t^{\mathrm{b}}$. We normalized both schemes with the performance of the periodic scheme under $\mathcal{I}_{t,i}^{\mathrm{bl}}$ given equal global average triggering rates according to (11). While the results for $\mathcal{I}_t^{\mathrm{b}}$, namely $J_{\mathrm{ET}}^{\mathrm{b}}(\Delta_{\mathrm{b}})/J_{\mathrm{TT}}^{\mathrm{bl}}(\mathbb{E}\big[T_{\mathrm{ET}}^{\mathrm{b}}(\Delta_{\mathrm{b}})\big]/n) = n/3$, can be obtained analytically, the results for $J_{\mathrm{ET}}^{\mathrm{bl}}(\Delta_{\mathrm{bl}})/J_{\mathrm{TT}}^{\mathrm{bl}}(\mathbb{E}\big[T_{\mathrm{ET}}^{\mathrm{bl}}(\Delta_{\mathrm{bl}})\big])$ are obtained numerically as described in [19, Sec. 5]. Thus, utilizing more information locally improves the performance of the ETC scheme.

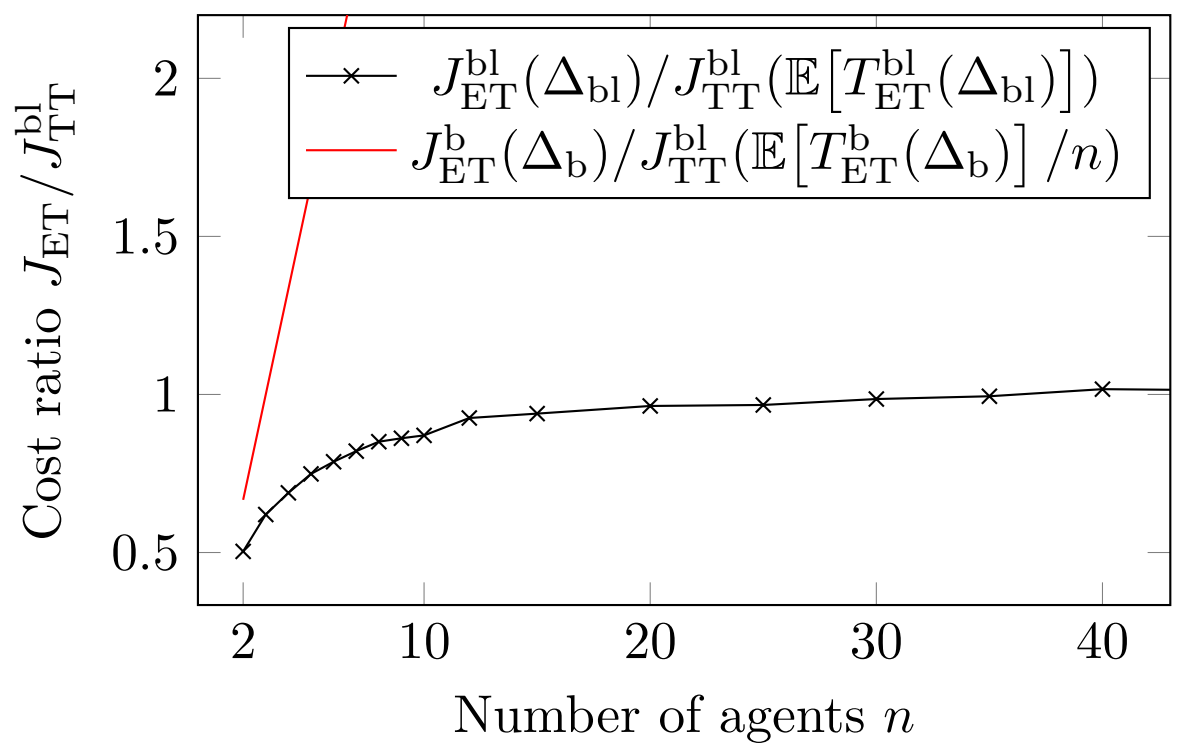

Fig. 1. Performance comparison of ETC schemes under $\mathcal{I}_t^{\mathrm{b}}$ and $\mathcal{I}_{t,i}^{\mathrm{bl}}$: We normalized both schemes by the performance of the periodic scheme under $\mathcal{I}_{t,i}^{\mathrm{bl}}$. We compare and normalize the schemes under equal average triggering rates, i.e., under constraint (11).

Secondly, we emphasize that it depends on the application which scheme is actually applicable. For example, [4] arrives at control input (4) uniquely via a stochastic leader problem. Thus, the exact choice of control inputs in Theorems 3 and 5 depends on the intended use case. In addition, the locally utilized information depends on the setup since, in certain applications, the information $\mathcal{I}_{t,i}^{\mathrm{bl}}$ may not be available whereas $\mathcal{I}_t^{\mathrm{b}}$ is, e.g., when local controller and sensor are separated and only communicate at events.

Thirdly, note that, apart from the locally utilized information, the settings analyzed in this paper are conceptually equivalent. In particular, both ETC schemes utilize level-triggering rules of the same form. Therefore, this work demonstrates that augmenting an information set with more information might render a previously consistent ETC scheme inconsistent while still improving the performance. We can observe this when moving from $\mathcal{I}_t^{\mathrm{b}}$ to $\mathcal{I}_{t,i}^{\mathrm{bl}}$ while utilizing the respective optimal controller.

Lastly, recall that [18] examines a setting for which the analysis under $\mathcal{I}_{t,i}^{\mathrm{bl}}$ applies whereas [4] proposes a scheme where the results under $\mathcal{I}_t^{\mathrm{b}}$ hold. On the one hand, [18] demonstrates that a natural extension of previously consistent ETC schemes to a distributed problem may become inconsistent. On the other hand, [4] shows that, in

Table 1
Simulation results for cost and average inter-event time

| $n$ | $J^{\mathrm{b}}_{\mathrm{TT}}$ | $J^{\mathrm{b}}_{\mathrm{ET}}$ | $J^{\mathrm{bl}}_{\mathrm{TT}}$ | $J^{\mathrm{bl}}_{\mathrm{ET}}$ |
|---|---|---|---|---|
| 3 | 2.273 (0.25 s) | 0.774 (0.261 s) | 0.748 | 0.457 (0.244 s) |
| 3 | 4.525 (0.50 s) | 1.558 (0.507 s) | 1.506 | 0.948 (0.505 s) |
| 10 | 222.2 (0.50 s) | 76.22 (0.514 s) | 22.55 | 20.53 (0.514 s) |
| 50 | 30223 (0.50 s) | 10375 (0.511 s) | 612.7 | 640.4 (0.516 s) |

For each cost entry, we also provide the global average inter-event times obtained from simulation in parantheses. The ones for $J^{\mathrm{bl}}_{\mathrm{TT}}$ are equivalent to column $J^{\mathrm{b}}_{\mathrm{TT}}$. Note that, in line with the explanations in Section 4.3, the global average inter-event times for the cases $J^{\mathrm{b}}_{\mathrm{TT}}$ and $J^{\mathrm{b}}_{\mathrm{ET}}$ are given by $T^{\mathrm{b}}_{\mathrm{TT}}/n$ and $\mathbb{E}\left[T^{\mathrm{b}}_{\mathrm{ET}}\right]/n$, respectively. To achieve the shown average inter-event times with ETC, we utilize $\mathbb{E}\left[T^{\mathrm{b}}_{\mathrm{ET}}(\Delta_{\mathrm{b}})\right] = \Delta_{\mathrm{b}}^2$. Moreover, for $\mathbb{E}\left[T^{\mathrm{bl}}_{\mathrm{ET}}(\Delta_{\mathrm{bl}})\right]$, we tune $\Delta_{\mathrm{bl}}$ in simulation to obtain the respective average inter-event times, resulting in $\Delta_{\mathrm{bl}} \in \{0.72, 1.04, 1.44, 1.90\}$ for the four rows.

a different distributed setting, this extension still results in a consistent scheme. While the analysis, schemes, and outcomes in [4] and [18] differ substantially, we unified the problem frameworks and generalized the results such that the relationship between the two works becomes apparent: they are not contradictory but complementary in the sense that they implicitly study consensus problems under different information assumptions. Thus, we have found the root of the different outcomes in [4] and [18]. We have thereby demonstrated that ETC results for single-loop setups can be transferred to distributed problems including consistency properties in some cases whereas this might be more challenging in other cases.

## 5 Simulation example

In this section, we simulate the MAS under the control inputs (4), (6) for different agent numbers and average triggering rates of TTC and ETC. Subsequently, we estimate the cost induced by the respective schemes based on these simulations. We utilize the Euler-Maruyama method with a step size of $2 \cdot 10^{-3}$ s over a time horizon of 2000 s. We deploy the triggering schemes and control inputs in simulation and observe the resulting closed-loop behavior on a sufficiently long time horizon.

We provide cost results for 4 different scenarios in Table 1, either differing in the number of agents $n$ or in the global average inter-event time shown in parantheses behind each cost value. Firstly, note that the cost ratios remain approximately constant for both scenarios with $n = 3$. This matches our theoretical insight that the ratio is not influenced by the choice of $\Delta$ or, more precisely, the choice of the global average inter-event time, see also [19]. Secondly, our findings from Theorem 9 and (12) are confirmed by the numerical results: The cost ratios $J^{\mathrm{b}}_{\mathrm{ET}}/J^{\mathrm{b}}_{\mathrm{TT}}$ and $J^{\mathrm{b}}_{\mathrm{TT}}/J^{\mathrm{bl}}_{\mathrm{TT}}$ are approximately $1/3$ and $n$, respectively, for all scenarios. Thirdly, note that for $n = 50$ the cost $J^{\mathrm{bl}}_{\mathrm{TT}}$ is lower than $J^{\mathrm{bl}}_{\mathrm{ET}}$ while this relationship is flipped for $n \in \{3, 10\}$. This demonstrates the loss of the consistency property for sufficiently large agent numbers $n$ for the ETC scheme with a controller utilizing broadcast and local information at triggering instants. Fourthly, we observe in all scenarios that the cost induced by the schemes only relying on broadcast information is larger than the one of schemes additionally utilizing local information at triggering instants. Consequently, utilizing richer information at the local controllers improves closed-loop performance significantly. However, consistency of the respective ETC scheme is lost for sufficiently large agent numbers if we move from schemes only relying on broadcast information to schemes additionally utilizing local information at triggering instants. The simulation results thus confirm our theoretical findings.

## 6 Conclusions

We have examined the performance properties of ETC compared to TTC for a continuous-time single-integrator consensus setup and a decentralized level-triggering rule. Note that we have derived optimal control inputs under the considered performance measure and triggering schemes. We have contrasted two information scenarios for the local controllers: one relying only on broadcast state information whereas the other one additionally utilizes local state information at triggering instants. We have shown that the former one renders ETC with a decentralized level-triggering rule consistent, i.e., more performant than TTC under the same average triggering rate, whereas the latter one yields the opposite result for sufficiently many agents.

The findings in this article demonstrate that the information utilized in the closed loop can have a decisive influence on the performance properties of ETC schemes, especially with respect to TTC performance. In particular, enriching the locally utilized information of the controller can render a consistent ETC scheme inconsistent while still improving closed-loop performance in general. Moreover, our results unravel the relationship between the contrasting consistency results of [4] and [18]: The difference in the locally utilized information explains the differing consistency properties of the otherwise similar ETC schemes.

Consequently, this work sheds light on the importance of locally utilized information for ETC performance. In future research, the findings from this very particular setup should be examined in more general settings with respect to system dynamics, communication topologies, performance measure, or triggering rules. While it remains open how the results generalize to other setups, the uncovered phenomenon can provide new insights into the performance properties of (decentralized) ETC and inspire new research on this topic in the future.

## A Proof of Theorem 3

Firstly, note that performance measure (2) is minimized if $\mathbb{E}\Big[\frac{1}{2}\sum_{(i,j)\in\mathcal{E}}(x_i(t)-x_j(t))^2\Big]$ is minimized for all $t$. Thus, let us minimize

$$\begin{aligned}
&\mathbb{E}\left[\frac{1}{2}\sum_{(i,j)\in\mathcal{E}}(x_i(t)-x_j(t))^2\right]\\
&=\frac{1}{2}\sum_{(i,j)\in\mathcal{E}}\mathbb{E}\Big[\mathbb{E}\Big[\Big(v_i(t)-v_i(t^i_{\hat{k}_i(t)})+\tilde{x}_i(t)\\
&\qquad -\ (v_j(t)-v_j(t^j_{\hat{k}_j(t)})+\tilde{x}_j(t))\Big)^2\Big|\mathcal{I}^{\mathrm{b}}_t\Big]\Big]\\
&=\frac{1}{2}\sum_{(i,j)\in\mathcal{E}}\Big(\mathbb{E}\Big[(v_i(t)-v_i(t^i_{\hat{k}_i(t)})-(v_j(t)-v_j(t^j_{\hat{k}_j(t)})))^2\Big]\\
&\qquad +\mathbb{E}\Big[\mathbb{E}\Big[(v_i(t)-v_i(t^i_{\hat{k}_i(t)})-(v_j(t)-v_j(t^j_{\hat{k}_j(t)})))\\
&\qquad\qquad \cdot\ (\tilde{x}_i(t)-\tilde{x}_j(t))|\mathcal{I}^{\mathrm{b}}_t]]\\
&\qquad +\mathbb{E}\big[(\tilde{x}_i(t)-\tilde{x}_j(t))^2\big]\Big), \qquad\qquad (A.1)
\end{aligned}$$

with the shorthand $\tilde{x}_i(t)=\sum_{k\in\mathbb{N}}H(t-t^i_k)(v_i(t^i_k)-v_i(t^i_{k-1}))+\int_0^t u_i(s)\,\mathrm{d}s$ for all $i\in\mathcal{V}$.

Secondly, note that $(t_k)_k$ and hence also $(t^i_k)_k$ for any $i\in\mathcal{V}$ are sequences of symmetric stopping times. Consequently, we have that, for all $i\in\mathcal{V}$, $\mathbb{E}[v_i(t)-v_i(t^i_{\hat{k}_i(t)})|\mathcal{I}^{\mathrm{b}}_t]=0$ as well as, for all $i,j\in\mathcal{V}$, $\mathbb{E}[(v_i(t)-v_i(t^i_{\hat{k}_i(t)}))(\sum_{k\in\mathbb{N}}H(t-t^j_k)(v_j(t^j_k)-v_j(t^j_{k-1})))|\mathcal{I}^{\mathrm{b}}_t]=0$. This is due to the fact that the distribution of the respective random variables is rendered symmetric for any $t$ due to the symmetry of the stopping times. Note that the expected value of random variables with symmetric probability distributions (centered around zero) is zero. Furthermore, as $v_i(t)-v_i(t^i_{\hat{k}_i(t)})$ and $u_j(s)$ are independent for any $i,j\in\mathcal{V}$ and $s\in[0,t]$, we can use these observations to establish $\mathbb{E}[(v_i(t)-v_i(t^i_{\hat{k}_i(t)}))\int_0^t u_j(s)\,\mathrm{d}s|\mathcal{I}^{\mathrm{b}}_t]=0$ for all $i,j\in\mathcal{V}$. Thus, the middle term in (A.1) is equal to zero.

Thirdly, due to the two remaining terms in (A.1) being quadratic, we can conclude that the cost is minimized by control inputs satisfying

$$\begin{aligned}
&\int_0^t u_i(s)\,\mathrm{d}s-\int_0^t u_j(s)\,\mathrm{d}s\\
&\qquad=\sum_{k\in\mathbb{N}}H(t-t^j_k)(v_j(t^j_k)-v_j(t^j_{k-1}))-(\star)_i
\end{aligned}$$

for all $(i,j)\in\mathcal{E}$ and $(\star)_i$ denoting the preceding expression but with $j$ replaced by $i$. This results from setting the last term in (A.1) to zero.

With this sufficient optimality condition at hand, we can now prove that the provided class of inputs is indeed optimal. For that, first note that, for the input provided in the theorem, $\int_0^t u_i(s)\,\mathrm{d}s=\sum_{k\in\mathbb{N}}H(t-t_k)\sigma_i(t)(c(\mathcal{I}^{\mathrm{b}}_t,t)-x_i(t))+\sum_{k\in\mathbb{N}}H(t-t_k)(1-\sigma_i(t))(c(\mathcal{I}^{\mathrm{b}}_t,t)-c(\mathcal{I}^{\mathrm{b}}_{t^-},t))$ where $\sigma_i(t)=1$ if $t=t^i_k$ for some $k\in\mathbb{N}_0$, and 0 otherwise. Hence,

$$\begin{aligned}
&\int_0^t u_i(s)\,\mathrm{d}s-\int_0^t u_j(s)\,\mathrm{d}s\\
&\qquad=\sum_{k\in\mathbb{N}}H(t-t^j_k)(x_j(t)-c(\mathcal{I}^{\mathrm{b}}_{t^-},t))-(\star)_i,
\end{aligned}$$

where $(\star)_i$ again denotes the preceding expression with $i$ instead of $j$.

To show $x_i(t^i_k)-c(\mathcal{I}^{\mathrm{b}}_{t^{i-}_k},t^i_k)=v_i(t^i_k)-v_i(t^i_{k-1})$, we note $c(\mathcal{I}^{\mathrm{b}}_{t^-_k},t)=c(\mathcal{I}^{\mathrm{b}}_{t_{k-1}},t)$ for $(t_k)_k$ and arbitrary $t$. We can thereby recursively deduce $c(\mathcal{I}^{\mathrm{b}}_{t^{i-}_k},t^i_k)=c(\mathcal{I}^{\mathrm{b}}_{t^i_{k-1}},t^i_k)+\sum_{\ell\in\mathbb{N}:t_\ell\in(t^i_{k-1},t^i_k)}(c(\mathcal{I}^{\mathrm{b}}_{t_l},t^i_k)-c(\mathcal{I}^{\mathrm{b}}_{t^-_l},t^i_k))$. With this fact and the integrated dynamics $x_i(t^i_k)=x_i(t^i_{k-1})+v_i(t^i_k)-v_i(t^i_{k-1})+\int_{t^i_{k-1}}^{t^i_k}u_i(s)\,\mathrm{d}s$, we arrive at

$$\begin{aligned}
&x_i(t^i_k)-c(\mathcal{I}^{\mathrm{b}}_{t^{i-}_k},t^i_k)\\
&=x_i(t^i_{k-1})+v_i(t^i_k)-v_i(t^i_{k-1})+\Big(c(\mathcal{I}^{\mathrm{b}}_{t^i_{k-1}},t^i_k)\\
&\qquad -x_i(t^i_{k-1})+\sum_{\substack{\ell\in\mathbb{N}:\\ t_\ell\in(t^i_{k-1},t^i_k)}}(c(\mathcal{I}^{\mathrm{b}}_{t_l},t^i_k)-c(\mathcal{I}^{\mathrm{b}}_{t^-_l},t^i_k))\Big)\\
&\quad -\Big(c(\mathcal{I}^{\mathrm{b}}_{t^i_{k-1}},t^i_k)+\sum_{\substack{\ell\in\mathbb{N}:\\ t_\ell\in(t^i_{k-1},t^i_k)}}(c(\mathcal{I}^{\mathrm{b}}_{t_l},t^i_k)-c(\mathcal{I}^{\mathrm{b}}_{t^-_l},t^i_k))\Big)\\
&=v_i(t^i_k)-v_i(t^i_{k-1}),
\end{aligned}$$

as desired.

## B Proof of Lemma 6

Let $Q(M):=\mathbb{E}\Big[\int_0^M x(t)^\top Lx(t)\,\mathrm{d}t\Big]$. Firstly, since an optimal control input according to Theorem 3 is deployed, we arrive at

$$\begin{aligned}
Q(M)&=\frac{1}{2}\sum_{(i,j)\in\mathcal{E}}\mathbb{E}\Big[\int_0^M(v_i(t)-v_i(t^i_{\hat{k}_i(t)})\\
&\qquad\qquad -\ (v_j(t)-v_j(t^j_{\hat{k}_j(t)})))^2\,\mathrm{d}t\Big]\\
&=(n-1)\sum_{i=1}^n\mathbb{E}\left[\int_0^M(v_i(t)-v_i(t^i_{\hat{k}_i(t)}))^2\,\mathrm{d}t\right].
\end{aligned}$$

Note that, in the last line, we have removed the expected value of the cross terms: Because of the symmetry of the stopping time, we have that, for $i \neq j$, the random variable $\int_0^M (v_i(t) - v_i(t^i_{\hat{k}_i(t)}))(v_j(t) - v_j(t^j_{\hat{k}_j(t)}))\,\mathrm{d}t$ has a symmetric distribution, yielding an expected value of 0.

Due to homogeneous agents and stopping rules, we have

$$\frac{Q(M)}{n(n-1)} = \Bigg(\mathbb{E}\Bigg[\sum_{k=1}^{m(M)} \int_{t^1_{k-1}}^{t^1_k} (v_1(t) - v_1(t^1_{k-1}))^2\,\mathrm{d}t\Bigg] + \mathbb{E}\Bigg[\int_{t^1_{m(M)}}^{M} (v_1(t) - v_1(t^1_{m(M)}))^2\,\mathrm{d}t\Bigg]\Bigg),$$

where $(m(M))_{M\in[0,\infty)}$ is the renewal process of the renewal time sequence $(t^1_k)_k$.

Since the inter-event times $(t^i_k - t^i_{k-1})_{k\in\mathbb{N}}$ are i.i.d., we have that $y^1_k := \int_{t^1_{k-1}}^{t^1_k} (v_1(t) - v_1(t^1_{k-1}))^2\,\mathrm{d}t$ are i.i.d. as well. Utilizing Wald's equation yields $\mathbb{E}\Big[\sum_{k=1}^{m(M)} y^1_k\Big] = \mathbb{E}[m(M)]\,\mathbb{E}\big[y^1_k\big]$. Moreover, the last cost component is bounded as $\int_{t^1_{m(M)}}^{M} (v_1(t) - v_1(t^1_{m(M)}))^2\,\mathrm{d}t \leq y^1_{m(M)+1}$. Consequently, dividing by $M$ and letting $M \to \infty$ yields

$$\begin{aligned} J &= n(n-1) \lim_{M\to\infty} \frac{\mathbb{E}[m(M)]}{M} \mathbb{E}\big[y^1_1\big] \\ &= n(n-1) \frac{1}{\mathbb{E}[T_1]} \mathbb{E}\Bigg[\int_0^{T_1} v_1(t)^2\,\mathrm{d}t\Bigg] =: J^{\mathrm{b}}(T_1), \end{aligned}$$

where we utilized that $\mathbb{E}[T_1] < \infty$.